\documentclass[conference]{IEEEtran}
\usepackage{graphicx}
\usepackage{balance}
	\usepackage{amsmath}
	\usepackage{amsfonts}
    \usepackage{longtable}
	\usepackage{fancyhdr}
    \usepackage{fancybox}
    \usepackage{booktabs}
	\usepackage{enumerate}
    \usepackage{framed}
    \usepackage{pgfplots}
	\usepackage{listings}
    \usepackage{subcaption}

\ifCLASSOPTIONcompsoc
  \usepackage[nocompress]{cite}
\else
  \usepackage{cite}
\fi
\ifCLASSINFOpdf
\else
\fi

\usepackage[framemethod=tikz]{mdframed}
\mdfdefinestyle{style1}{leftmargin=1cm,rightmargin=0.5cm,skipbelow=0.5cm,skipabove=0.3cm,innertopmargin=4pt}

\usepackage{hhline}
\usepackage{dcolumn}
\usepackage[utf8]{inputenc}
\usepackage{amsmath}
\newcolumntype{d}[1]{D{,}{.}{#1}}

\begin{document}
%
\title{Combinatorial Modeling and Test Case Generation for Industrial Control Software using ACTS}
%
%
%
%

\author{Sara Ericsson,
        Eduard Enoiu, \\
        M\"alardalen University, V\"aster{\aa}s,  Sweden.
\thanks{Manuscript received April 19, 2005; revised August 26, 2015.}}

%
%

\markboth{Journal of \LaTeX\ Class Files,~Vol.~14, No.~8, August~2015}%
{Shell \MakeLowercase{\textit{et al.}}: Bare Advanced Demo of IEEEtran.cls for IEEE Computer Society Journals}
%



\IEEEtitleabstractindextext{%
\begin{abstract}
Combinatorial testing has been suggested as an effective method of creating test cases at a lower cost. However, industrially applicable tools for modeling and combinatorial test generation are still scarce. As a direct effect, combinatorial testing has only seen a limited uptake in industry that calls into question its practical usefulness. This lack of evidence is especially troublesome if we consider the use of combinatorial test generation for industrial safety-critical control software, such as are found in trains, airplanes, and power plants.

To study the industrial application of combinatorial testing, we evaluated ACTS, a popular tool for combinatorial modeling and test generation, in terms of applicability and test efficiency on industrial-sized IEC 61131-3 industrial control software running on Programmable Logic Controllers (PLC). We assessed ACTS in terms of its direct applicability in combinatorial modeling of IEC 61131-3 industrial software and the efficiency of ACTS in terms of generation time and test suite size. We used 17 industrial control programs provided by Bombardier Transportation Sweden AB and used in a train control management system. Our results show that not all combinations of algorithms and interaction strengths could generate a test suite within a realistic cut-off time. The results of the modeling process and the efficiency evaluation of ACTS are useful for practitioners considering to use combinatorial testing for industrial control software as well as for researchers trying to improve the use of such combinatorial testing techniques. 

\end{abstract}

}

\maketitle

\IEEEdisplaynontitleabstractindextext

%
\IEEEpeerreviewmaketitle

\ifCLASSOPTIONcompsoc
\IEEEraisesectionheading{\section{Introduction}\label{sec:introduction}}
\else
\section{Introduction}
\label{sec:introduction}
\fi
Software testing is an important activity used during software development \cite{ammann2008introduction} for quality assurance. A test case used during testing includes input parameters for stimulating the software under test (i.e., a sequence of input parameters) and the expected behavior in terms of outputs. Engineers are designing a set of test cases called a test suite and are using a test framework for running the test suite against the software under test. The execution of the test suite produces a test result which is compared to the expected result as stated in the requirements. The result of executing a test suite is either a pass or a fail verdict. Traditionally, the creation of tests is performed manually by human engineers or supported by an automatic test case generation tool able to design test cases based on certain criteria. In industrial practice, because of the complexity of the created software \cite{nie2011survey} there is no practical way to exhaustively test all the possible combinations between input parameter values. The selection of efficient and effective test cases has been a topic of research for the last couple of decades \cite{ammann2008introduction}. In particular, executing all possible input combinations is often not practical since not all combinations of parameter values trigger an interesting behavior and help in the detection of a software fault. Combinatorial testing has been proposed as a suitable method for efficient and effective automatic test design \cite{nie2011survey,kuhn2010sp}. Combinatorial testing has been an active field of research for more than 20 years \cite{nie2011survey} with a number of combinatorial test generation tools being proposed and developed. Advanced Combinatorial Testing System  (ACTS) \cite{yu2013acts,acts292userguide17} is one of the most popular combinatorial test generation tools used in software testing research. Even if ACTS has been evaluated on several case studies \cite{kuhn2010sp}, there is a lack of evidence on how combinatorial testing tools can be used for modeling and testing of industrial control software, like the one used in the safety-critical domain. In this domain, faults can lead to economical damage and, in some cases, loss of human lives. As a consequence there is a need to investigate the use of combinatorial testing and identify the empirical evidence for, or against, modeling and test generation in practice when developing industrial control software.

This study shows how ACTS can be used for input space modeling and evaluates the efficiency of the test generation process when applied to industrial control software. The study strives to answer how ACTS can be used for test generation for a given set of programs provided by Bombardier Transportation, a large company manufacturing train software and hardware. The programs are developed in the IEC 61131-3 \cite{international3iec} programming standard for process control software and Programmable Logic Controllers (PLCs), commonly used in the engineering of embedded safety-critical systems (e.g., in the railway and power control domains). The programs are used in a train control management system (TCMS) running on PLCs. The TCMS is in charge of many of the safety related control and operational-critical functionality of a train, which makes it crucial to detect software faults as early as possible and in an efficient way. The use of automatic test generation using ACTS could help an industrial engineer in testing the PLC software in a more efficient way. Therefore, it is relevant to evaluate the applicability of modeling the PLC programs and to study how efficient the test generation for PLC software is when using ACTS.

The aim of this paper is to evaluate combinatorial testing on industrial control software and how it can be used in practice for modeling and test generation for PLC software. In addition, we provide valuable insights and experiences in how to apply combinatorial testing in industrial practice. 
To achieve the mentioned goal, we have designed an
experiment to answer the following research questions:
\begin{itemize}
\item \textit{\textbf{RQ1}: Is combinatorial testing using ACTS applicable for modeling industrial PLC software?} This question intends to assess the
capability of ACTS which is a widely adopted tool in academia, in terms of its modeling capabilities for real industrial control software.
\item \textit{\textbf{RQ2}: How efficient is combinatorial testing using ACTS for test generation of industrial control software?} This question
aims to evaluate the test generation efficiency of different combinatorial test criteria.  
\end{itemize}

To answer these questions we applied ACTS to 17 industrial PLC programs of different sizes. We manually modeled the input space and automatically generated test suites using ACTS. In addition, we evaluated the test generation capability in terms of efficiency and applicability based on the generation time and the number of obtained test cases.

\section{Background}
Testing is an important part of software development. As exhaustive testing is costly and practically impossible to use when dealing with industrial software \cite{kuhn2010sp} there is a need for applicable test generation methods. There is some compelling experimental evidence showing that most faults are triggered by a combination of a few parameter values \cite{kuhn2010sp}, which makes combinatorial testing a feasible method for creating efficient and effective test cases.

\subsection{Combinatorial Testing}
Combinatorial testing can be helpful in the creation of test cases by generating certain combinations among parameter values. Several techniques have been proposed for combinatorial testing \cite{kuhn2009combinatorial,nie2011survey,kuhn2010sp,bryce2010combinatorial} in order to generate a test suite which covers the combinations of $T$ parameter values at least once or in a certain interaction strategy (e.g., each-used, pair-wise, t-wise, base choice). One of the most used combinatorial testing criterion is pairwise or 2-wise. A test set generated using pairwise covers all the possible pairs of parameters in the input space \cite{kuhn2009combinatorial,kuhn2008practical,tai2002test}. There are a few empirical studies showing that the majority of software faults (i.e., between 50 to 97\% according to a study by Kuhn et al. \cite{kuhn2009combinatorial} from 2009, and 60 to 90\% according to another study \cite{kuhn2010sp} from 2010) can be discovered using pairwise testing or a higher criterion. This is a promising result that motivates the need to study such techniques in industrial projects. The results of another study \cite{kuhn2008practical} showed that most of the faults have been discovered by test suites created using less than seven parameter combinations \cite{kuhn2009combinatorial}. Since there is some evidence showing that software failures are triggered by a combination of 6 or fewer input parameter values \cite{kuhn2010sp,bartholomew2013industry}, we are interested in investigating the applicability and scalability of automatically creating test cases using these combinatorial techniques for industrial programs. By using higher strength t-way testing in the detriment of manually created test cases can directly influence the cost-effectiveness and applicability of such techniques. 
Several testing strategies have been proposed to reduce the number of combinations that have to be enumerated and the time to generate test cases. 
A combinatorial testing strategy called In-Parameter-Order (IPO) has been used in several different algorithms to handle problems such as the sheer number of test cases and long generation times \cite{gao2014general} for 2-way testing. The In-Parameter-Order-General (IPOG) algorithm has been proposed as a variant of the IPO strategy that supports higher strength than 2-way \cite{lei2007ipog,lei2008ipog}. The IPOG-D strategy is an extension of IPOG combined with a recursive approach instead of enumerating combinations used by IPOG, making it suitable for larger multi-way combinations \cite{lei2008ipog}.

Modeling the software under test is an important part of using combinatorial testing since it affects the test generation procedure \cite{nie2011survey,borazjany2012combinatorial}. According to Nie and Leung \cite{nie2011survey} a combinatorial test model for the software under test consists of the following four elements: parameters, values of parameters, possible interaction relations among parameters and parameter constraints used to exclude combinations that are not possible. A model consisting of these elements can be created by reviewing different system documents and the code of the program, along with other artifacts. 
We are interested, in this paper, to investigate the creation of the test model for real-world industrial control software and to investigate what the sources of information used for creating these modeling elements are.

\subsection{ACTS Tool}
ACTS is a combinatorial test generation research tool, previously known as FireEye \cite{lei2007ipog}. It is an academic tool developed by researchers at the National Institute of Standards and Technology and the University of Texas at Arlington in the United States. It was first released in 2006 \cite{yu2013acts} and seems to be actively updated. ACTS supports t-way combinatorial test generation, with t restricted to values between one and six. The tool has been evaluated and has been shown to potentially uncover the faults caused by the combination of up to six parameters \cite{yu2013acts}. A set of parameters is defined as a system under test (SUT) and the tool supports the following data types: Boolean, Number, Enum and Range \cite{acts292userguide17}. A number of algorithms for combinatorial test generation are implemented in ACTS version 2.92: \textit{IPOG }\cite{lei2007ipog}, \textit{IPOG-F, IPOG-F2, IPOG-D }and \textit{Base Choice}. In base choice criterion, the value of one input at the time is varied while keeping all other inputs at fixed base values until all combinations have been covered. This process is repeated for each parameter to create a final test suite. According to the user guide of ACTS, IPOG-D is the preferred strategy for larger systems while the other algorithms are better suited for moderate-sized systems, with less than 20 parameters and 10 values per parameter on average \cite{acts292userguide17}. In this study, we evaluate this hypothesis by using several programs of different sizes.

ACTS also supports the following two different test generation modes: scratch and extend. The former creates a whole new test set, while the latter extends an existing test suite with automatically generated tests. ACTS supports, as previously mentioned, t-way test generation and mixed combination strengths (i.e., different values of t) for grouped parameters. 
A set of parameters with the same strength is defined as a relation. Another feature in ACTS is the possibility to add constraints, since some combinations might not be allowed and should therefore not be a part of the test suite. The user can define constraints and the test suite will be generated with test cases which fulfill these constraints. Once a test suite has been generated it is possible to verify the t-way coverage, i.e. how many of the tests cover the t-way coverage criteria \cite{yu2013acts,acts292userguide17}. ACTS is platform independent and implemented in Java programming language. The tool can be used in several ways since ACTS provides a graphical user interface, an application-programming and a command line interface \cite{yu2013acts}. The test suite can be saved to a file and exported in a comma separated value format \cite{acts292userguide17}.

Other combinatorial test generation tools have been proposed and are available. The IPO (PairTest) \cite{tai2002test} and the AETG System \cite{cohen1997aetg} are also used for pairwise testing implementing the IPO strategy. A list of available tools for pairwise testing can be found in \cite{czerwonka2009pairwise}. In this study we use ACTS, as this is a mature tool that has been used in several industrial scenarios and is expected to be representative in the use of combinatorial testing techniques. 

\subsection{PLC Control Software}
Programmable Logic Controllers (PLCs) are a particular kind of computers used in automation systems and are widely used nowadays in the safety-critical domain \cite{bolton2015programmable,pavlovic2010model}. Automation tasks for industrial processes, such as automated packaging \cite{bolton2015programmable}, household appliances \cite{pavlovic2010model} and train control and management \cite{enoiu2016automated} are just some of the examples where PLCs are used. A PLC contains a microprocessor, a programmable memory and an I/O for communication \cite{bolton2015programmable}. The PLC runs in a cyclic loop according to a certain task level such that the program continuously reads the input parameters, executes the program and writes the output in every execution of the task \cite{pavlovic2010model}. An international standard has been proposed for programming PLCs named IEC 61131-3 \cite{international3iec}, published in 1993. Its purpose is to avoid the use of proprietary and closed versions of programming languages that cannot be used for other system applications and hardware. The standard defines the following five programming languages: Instruction List (IL) and Structured Text (ST) are textual languages while Ladder Diagram (LAD), Sequential Function Chart (SFC) and Function Block Diagram (FBD) are graphical languages. The IEC 61131 standard includes the description of these five languages and a complete development cycle that can be used for engineering software on this kind of a platform \cite{bolton2015programmable}. In this study we used programs developed in the IEC 61131-3 standard by industrial engineers, describing a safety-critical system used in the train domain. 

\begin{table*}[!t]
\renewcommand{\arraystretch}{1.3}
\caption{Program categories based on the information provided by three industrial engineers. Three categories have been defined: small, medium and large programs in terms of size and complexity.}
\label{table:engineer_input}
\centering
\begin{tabular}{|l|l|l|l|}
\hline
 & Small PLC Program & Medium PLC Program & Large PLC Program \\ \hline
Engineer 1 & 1-5 inputs & 5-10 inputs & 10-30 inputs \\ \hline
Engineer 2 & 1-7 inputs & 7-10 inputs & 10-20 inputs \\ \hline
Engineer 3 & 1-5 inputs & 5-10 inputs & 10-20 inputs \\ \hline
Average & 1-6 inputs & 6-10 inputs & 10-23 inputs \\ \hline   
\end{tabular}
\end{table*}

\subsection{Related Work}
Recently, researchers have shown an increased interest in combinatorial software testing. There are a number of studies in which combinatorial testing tools and techniques are being evaluated (e.g., \cite{borazjany2012combinatorial,hagar2015introducing,lei2007ipog}) in their use of combinatorial modeling and testing of industrial systems.  
Borazjany et al. \cite{borazjany2012combinatorial} performed a case study in which they applied combinatorial testing on an industrial system by using ACTS to generate tests. The purpose of their study was to apply combinatorial testing to a real-world system and evaluate its effectiveness, as well as gaining experience and insight in the practical application of combinatorial testing, including the input modeling process. The tests are generated for testing both the functionality and the system robustness. Another study conducted at Lockheed Martin \cite{hagar2015introducing} reports about an introduction of combinatorial testing in a large organization. The applicability of combinatorial testing was evaluated by comparing different features contained in a set of  combinatorial test tools and then applying these tools to real world systems. A number of pilot projects were conducted where ACTS was used as the primary tool. According to the results of this study, ACTS continued to be used by a number of teams once the pilot projects ended. Lei et al. \cite{lei2007ipog} conducted a study to generalize the pairwise IPO strategy to t-way testing, and implemented this new strategy in FireEye. The tool was evaluated in terms of efficiency by using different system configurations. The experiments showed that the number of tests increased rapidly with the t-strength. FireEye and four other test generation tools were applied to a Traffic Collision Avoidance System (TCAS) module. The results show that FireEye performed considerably better in both size of test suites and generation time for higher strength t-way testing. 

Evaluations of the applications of such combinatorial techniques on industrial systems are still rare, possibly because they require a level of efficiency and modeling maturity of the underlying tools that is difficult to achieve in academic tools. Consequently, there is a need to provide further evidence of the capabilities of modeling and test efficiency of automated combinatorial test generation on industrial systems. Another shortcoming of these empirical
studies on combinatorial test generation tools is that they are not focusing on the sources of information needed for modeling the input model and how applicable such techniques are for specific industrial systems, such as PLCs. Previous studies on industrial systems have shown that test generation tools are quite effective at detecting faults, but how efficient and applicable are
they for industrial application of different sizes and number of parameters? In this paper, we investigate this question using programs from an industrial train control management system. These studies motivated us to perform this study and provide some valuable steps towards the application of ACTS in industrial practice for PLC software.

\section{Experiment Setup}
In this study we have conducted a case study \cite{runeson2009guidelines} in which we applied combinatorial modeling and test generation to PLC software and evaluated the applicability of using such a technique on different industrial programs. The purpose of using ACTS for several PLC programs is to show its applicability and  test generation efficiency. In addition, we show how combinatorial testing can be used for domain specific programs, such as the ones used in PLCs. Once the modeling was performed, we generated test suites using different t-strengths and test generation strategies.

\begin{figure}[ht]
\includegraphics[width=\linewidth]{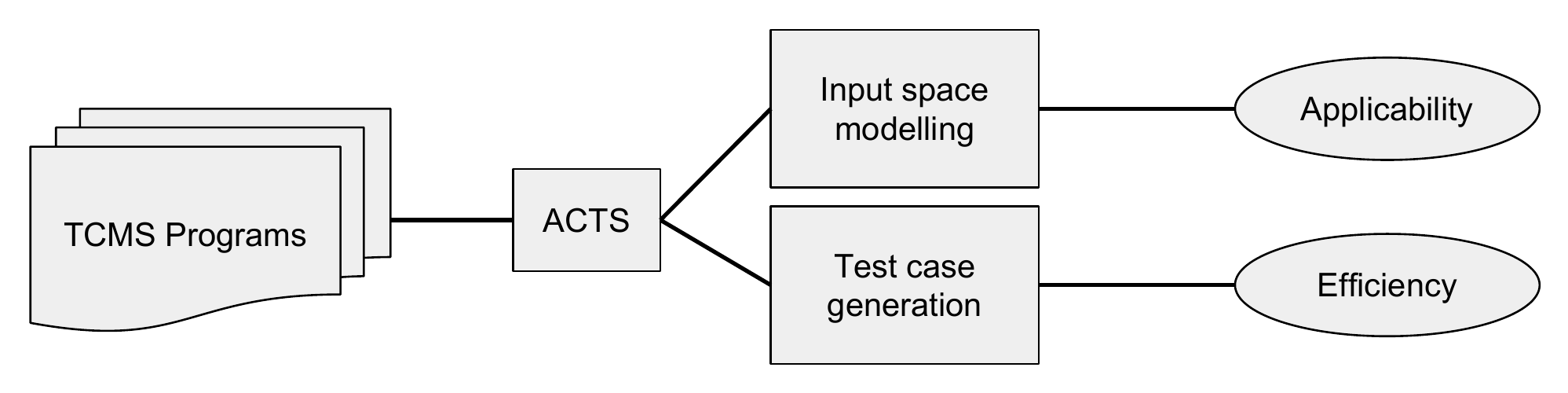}
\caption[Methodology]{Method for modelling and test generation using ACTS.}
\label{figure:methodology}
\end{figure}

The overall method of this study is shown in Figure \ref{figure:methodology}. PLC programs are used for input space modeling and test generation in ACTS. The input space modeling is performed manually in ACTS while the tests are generated automatically. The modeling step is evaluated in terms of applicability 
(i.e, can the selected programs be completely modeled using ACTS) 
and the test suite generation is evaluated in terms of efficiency 
(i.e., generation time and number of tests).

\subsection{Subject of Study}
The process began by selecting a number of representative PLC programs of different sizes to use for modeling and test generation using ACTS. 
We used programs from a control train system already developed by Bombardier Transportation Sweden AB. Due to time and resource constraints, 
we selected a subset of programs to be used for modeling using ACTS, from the provided Train Control Management System (TCMS) containing programs written in IEC 61131-3 FBD language. TCMS is a complex system with operation-critical functionality for which testing is very important in order to be able to achieve a safe and correct implementation in an efficient and effective manner. The software part of TCMS is part of a distributed system running on multiple hardware units distributed across a train.

\begin{figure}
\centering
\includegraphics[width=1.2\columnwidth]{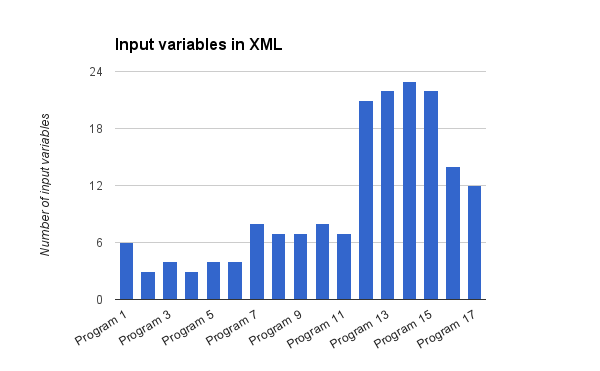}
\caption{The distribution of the number of input parameters per selected PLC program.}
\label{figure:input_per_program}
\end{figure}

Three engineers working at Bombardier Transportation Sweden AB and developing this system were asked to define three size and complexity categories for the PLC programs in TCMS: small, medium and large programs. Based on these categories we asked them to write down the corresponding number of inputs interval for each category according to their experience. The result of this data collection is shown in Table \ref{table:engineer_input}. In addition, we asked each engineer to select two programs from each category that contains typical and diverse behavior and data types found in PLC programs written in IEC 61131-3 FBD language. The engineers selected 18 programs in total out of the 189 PLC programs originally available. In determining the size of a program, the engineers took into account both the number of parameters and the number of values for each parameter. One program was removed from this experiment since it was categorized wrongly as a medium size program (i.e, this program should have been categorized as a small program). Finally, 17 programs were used for combinatorial modeling and testing (i.e. six small, five medium and six large programs). As shown in Figure \ref{figure:input_per_program}, the first six programs are small in size according to the categories defined by the industrial engineers. We considered a program with six inputs to be of small size. In addition, programs 7-11 are of medium size and programs 12-17 are large in size.

\subsection{Test Generation}
Once the models have been created using ACTS, test suites were generated for each model. ACTS provides the following algorithms for generating tests: IPOG, IPOG-F, IPOG-F2, IPOG-D and Base Choice. When using IPOG or IPOG-F the strength can be set to 1-6, while IPOG-F2 and IPOG-D support strength 2-6 for t-way coverage. For Base Choice the choice values can be assigned to each parameter. A base choice value can been seen as a typical or interesting value of the parameter. In this study a base value was assigned to each parameter during the modeling process by a test engineer experienced with testing PLC programs.

The tests were generated from the input space models and the systems created using ACTS, by running ACTS from the command-line. A batch file was created to generate several test suites from several programs consecutively. The output of the ACTS tool contained the relevant information, such as number of tests and generation time. This information was written to a log file. The creation of test suites has been performed on the same machine running on 8GB of RAM with 2.9GHz CPU. We used a batch file calling the ACTS application with the following parameters during the test generation: \textit{Ddoi} is the strength for t-way testing, \textit{Dalgo} is the used algorithm, \textit{Doutput} is the type of file the generated tests should be saved to, and \textit{Xmx8g} is used to increase the java heap space for the application to 8GB. All other options were automatically set to default in the ACTS command-line application when generating tests. The strength and the algorithm varied throughout the test generation and the generated test suites were saved as comma-separated values files. Test cases are generated by applying all possible algorithms on each program. We used different interaction strengths for all algorithms except BaseChoice and the test generation process was stopped when any of the following cases occurs:				 \\\\
1.\textit{ Strength is greater than the number of inputs.}	\\
2.\textit{ Maximum strength is reached.}							\\
3. \textit{Test generation time is greater than a certain time limit (i.e., 1 hour time limit has been used in this study after discussion with engineers in the company).}		\\
4.\textit{ ACTS application runs out of memory (heap space).}\\

When case (1) or case (2) occur no further test suites with higher strength can be generated, as the strength of the algorithm cannot be greater than the number of parameters and none of the algorithms provided by ACTS support a greater strength than six. The cut-off time for generating test cases (3) was set to one hour based on discussion with engineers in Bombardier Transportation Sweden AB testing the programs considered in this case study. No test was generated further for the remaining strengths when case (3) occurred because the measured test generation time from our initial experiments showed that time increases significantly with the strength of t. In addition, if the test cases are not generated within the given time limit, the tool is not showing the number of generated test cases. This information is not available and was not used when evaluating the efficiency in terms of number of test cases created. 

\begin{table}[!t]
\renewcommand{\arraystretch}{1.3}
\caption{The average number of values per parameter and the number of parameters per model.}
\label{table:param_values_avg}
\centering
\begin{tabular}{|l|d{6}|d{3}|}
\hline
\textbf{Program} & \textbf{Average} & \textbf{Parameters} \\ \hline
Program 1		&18,8		&6	\\ \hline
Program 2		&3,7		&3	\\ \hline
Program 3		&3,7		&4	\\ \hline
Program 4		&4,0			&3	\\ \hline
Program 5		&3,7		&4	\\ \hline
Program 6		&2,0			&34	\\ \hline
Program 7		&18,9		&8	\\ \hline
Program 8		&3,2		&7	\\ \hline
Program 9		&19,4		&7	\\ \hline
Program 10		&34,4		&8 	\\ \hline 
Program 11		&2,9		&7	\\ \hline
Program 12		&10,6		&21	\\ \hline 
Program 13		&7,7		&22	\\ \hline 
Program 14		&4,0			&23	\\ \hline	
Program 15		&19,4		&22	\\ \hline 
Program 16		&36,0			&14	\\ \hline 
Program 17		&13,6		&27	\\ \hline 
\end{tabular}
\end{table}

In the ACTS guide \cite{acts292userguide17}  the authors are recommending the use of IPOG, IPOG-F and IPOG-F2 for programs of moderate size, while using IPOG-D for larger programs. They define moderate programs in size as models with less than 20 parameters and approximately 10 values per parameter. Table \ref{table:param_values_avg} shows the number of parameters and the average amount of values per parameter for each modeled program. We evaluate the ACTS guide recommendation in the results section.

\subsection{Measuring Modeling Applicability}
The subject programs were analyzed by observing their IEC 61131-3 representation in an XML format. We checked if the information needed for input space modeling is included in the source code of the XML files. The information which could not be obtained directly from these files was gathered directly by informal interviews with engineers developing these programs at Bombardier Transportation Sweden AB. All information obtained using this collection of information was used to manually model the input space in ACTS. The input space modeling is evaluated in terms of applicability. Applicability refers to its success in meeting the input space modeling goal (i.e, can the selected programs be modeled in their entirety using ACTS and what are the sources of information needed). This information is key to determine the value of ACTS and its feasibility in an industrial context.

\subsection{Measuring Test Generation Efficiency and Cost}
The test cases were automatically generated using the ACTS tool. We collected the information needed to evaluate the test generation efficiency and cost (i.e., generation time and number of test cases).
The test case generation is evaluated in terms of efficiency by measuring the generation time (i.e. the time ACTS takes to generate a test suite) and the number of test cases generated by ACTS for each strategy and strength. These measures were collected directly as an output of running the ACTS tool on each program. 

We considered that a test suite is created efficiently if it has been generated in less than ten minutes. This specific time has been collected through interviews with three engineers working at Bombardier Transportation Sweden AB. According to this information, generating tests in less than 10 minutes is equivalent to an efficient manual testing session for creating specification-based tests for these programs.

To accurately measure the testing cost effort one would need to measure the costs for performing all testing activities. However, since the case study is focusing on test generation efficiency for a system that is in use and for which the development is finished, this type of cost measurement was not feasible. Instead, we collected the number of test cases generated as a proxy measure for the cost of testing. We are interested in investigating the cost of using different strategies in the same context. In this case study, we consider that the costs are related to the number of test cases. The higher the number of tests cases, the higher is the respective test suite cost. For example, testing a complex program will require more effort than a simple program. Thus, we assume that the cost measure is related to the complexity of the software which will influence the number of generated test cases. 
 
\section{Results}
The following section shows the result of modeling the PLC programs and the creation of an ACTS system for each program, followed by the generation of test suites and the collection of the results. For more details on the raw results obtained from this study we refer the reader to the dataset report \cite{saraeduard2018}. 

\subsection{Input Space Modeling}



\begin{figure}
\centering
\includegraphics[width=\columnwidth]{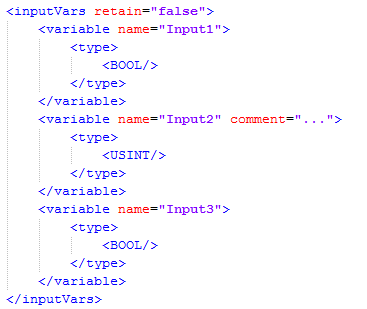}
\caption{A snippet of an XML file showing three input variables for a PLC control program. The comment option is one of the sources of information that can be used for modeling the input space.}
\label{figure:code_xml_program}
\end{figure}

\begin{figure*}
	\centering
	\begin{subfigure}[b]{0.50\textwidth}
		\includegraphics[width=\textwidth]					{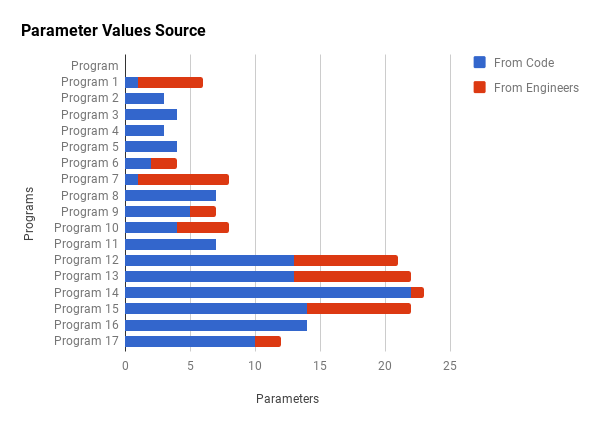}
		\caption{Parameter Values Source}
		\label{figure:diagram_paramvalues_src}
    \end{subfigure}%
         \hfill
    \begin{subfigure}[b]{0.50\textwidth}
    	\includegraphics[width=\textwidth]					{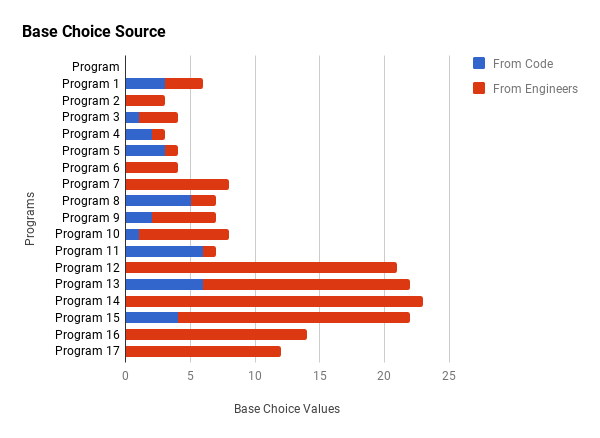}
        \caption{Base Choice Values Source}
        \label{figure:diagram_bc_source}
          \end{subfigure}
\caption{Distribution of the source of information for parameter values for each program and the source of information for base choice values for each program.}\label{fig:AllBoxes}
\end{figure*}

A model is defined by its parameters, the parameter constraints and possible relations among parameters. Each parameter is defined by a name, a data type and all the possible or permitted values \cite{nie2011survey}. In this study a base choice value is also assigned to each parameter. These base choice values are used when generating tests using the base choice strategy. The name and data type of each parameter were determined by manually analyzing the PLC programs provided as XML files, by parsing trough the code and the related comments. The possible parameter values for some of the parameters were determined by analyzing the XML files. These values were found by inspecting the commented code in the variable-tag for some of the input variables (as shown in the snippet displayed in Figure \ref{figure:code_xml_program}). 
All variables of boolean type were assigned their possible values (i.e., true and false). 
The rest of the information needed to model the input space was obtained by interviewing the engineers working at Bombardier Transportation Sweden AB developing and testing these programs. The engineers writing the code were asked to fill in the missing information, which in most cases provided the rest of the parameter values. The rest of the missing values and base choice values were obtained by asking a test engineer at Bombardier Transportation, responsible for testing some of the programs used in this case study. No constraints or relations could be defined for the PLC programs included in this study. Overall, we found out that all information needed for defining the ACTS model was either directly obtained from the code (i.e. by analyzing the XML files) or by interviewing the engineers writing and testing these programs. Figure \ref{figure:diagram_paramvalues_src} and Figure \ref{figure:diagram_bc_source} show the distribution of the sources from which the information needed for the input space modeling was obtained.

\begin{figure}
\centering
\includegraphics[width=1.2\linewidth]{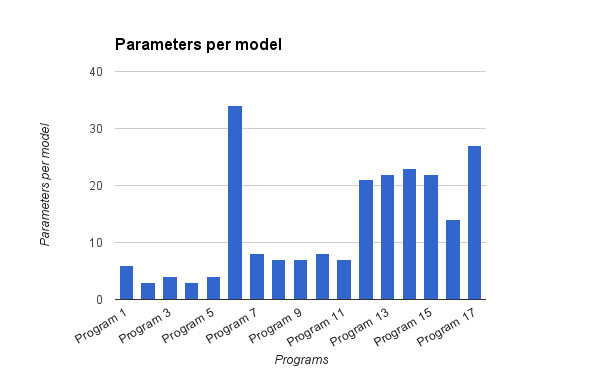}
\caption[Number of parameters for each modeled program]{Number of parameters for each modeled program.}
\label{figure:diagram_param_model}
\end{figure}

Two different kind of input space models were created for these programs, depending on the data types found in the programs. A concrete model refers to a representation in which tests generated based on this model can be directly used for testing. Abstract models differ from the concrete ones because they can not be used for testing without the use of an adapter for concretization. Hence, we first need to generate a set of abstract test cases. Some of the generated abstract tests need to be combined and enriched with glue information to form actual test cases \cite{ghandehari2013applying}. Figure \ref{figure:diagram_param_model} shows how many parameters each modeled program contains.
\begin{table}[!t]
\renewcommand{\arraystretch}{1.3}
\caption{The different data types and the number of occurrences of these types among the studied IEC 61131-3 programs.}
\label{table:types_xml_number}
\centering
\begin{tabular}{|l|d{3}|}
\hline
\textbf{Data Type in IEC 61131-3} & \textbf{Parameters} \\ \hline
BOOL & 79 \\ \hline
USINT & 45 \\ \hline
INT & 17 \\ \hline
UINT & 12 \\ \hline
REAL & 10 \\ \hline
UDINT & 5 \\ \hline
STRING & 4 \\ \hline
WORD & 3 \\ \hline
\end{tabular}
\end{table}
The different data types found in the PLC programs are shown in Table \ref{table:types_xml_number}. A concrete model was applied to 15 out of the 17 programs, while the remaining two used an abstract model. These two programs contained inputs of type WORD (i.e., 16 bit WORD data type). The models for these programs are abstract in the sense that each WORD is represented as 16 boolean parameters. Each parameter represents a bit in the WORD, and therefore the values of these parameters combined will form a complete WORD.

\begin{table}[!t]
\renewcommand{\arraystretch}{1.3}
\caption{Translation rules for different data types from IEC 61131-3 and their respective type in the ACTS model.}
\label{table:types_convert}
\centering
\begin{tabular}{|l|l|l|}
\hline
\textbf{Translation Rules} & \textbf{IEC 61131-3} & \textbf{ACTS} \\ \hline
Rule 1 & BOOL & Boolean \\ \hline
Rule 2 & INT & Number \\ \hline
Rule 3 & USINT & Number \\ \hline
Rule 4 & UINT & Number \\ \hline
Rule 5 & UDINT & Number \\ \hline
Rule 6 & REAL & Enum \\ \hline
Rule 7 & STRING & Enum \\ \hline
Rule 8 & WORD & Boolean \\ \hline
\end{tabular}
\end{table}

\begin{table}[!t]
\renewcommand{\arraystretch}{1.3}
\caption{The different data types and the number of occurrences of these types among the modeled ACTS systems.}
\label{table:types_modelling_number}
\centering
\begin{tabular}{|l|d{3}|}
\hline
\textbf{Data Type in ACTS} & \textbf{Parameters} \\ \hline
Boolean & 127 \\ \hline
Number & 79 \\ \hline
Enum & 14 \\ \hline
\end{tabular}
\end{table}

When the names, data types, possible parameter values and the base choice values are determined for each program, ACTS is used to model the input space. The ACTS GUI application is used for modeling and for each program a system was created. The types of parameters in the PLC code had to be converted to a corresponding data type in ACTS. The ACTS tool \cite{acts292userguide17} provides the following four data types: \textit{Boolean}, \textit{Number}, \textit{Range} and \textit{Enum}. Table \ref{table:types_convert} shows the data types found in the programs along with their corresponding type in ACTS. Range is considered the same data type as Number, since a Range consists of ranges of type Number.
In ACTS the Number and Range data types only support integer values. Therefore, the type Enum was used for floating point values. The total number of different data types among the modeled programs are shown in Table \ref{table:types_modelling_number}. 

\begin{table}[!t]
\renewcommand{\arraystretch}{1.3}
\caption{The number of parameter values of different sizes for every model before the reduction (VPP=Values per parameter).}
\label{table:param_values_before_reduction}
\centering
\begin{tabular}{|l|d{1}|d{1}|d{1}|}
\hline
Program&
VPP >99&
VPP (21-99)&
VPP <21 \\ \hline
Program 1	&2	&0	&4\\ \hline
Program 2	&0	&0	&3\\ \hline
Program 3	&0	&0	&4\\ \hline
Program 4	&0	&0	&3\\ \hline
Program 5	&0	&0	&4\\ \hline
Program 6	&0	&0	&34\\ \hline
Program 7	&1	&0	&7\\ \hline
Program 8	&0	&0	&7\\ \hline
Program 9	&2	&1	&4\\ \hline
Program 10	&2	&3	&3\\ \hline
Program 11	&0	&0	&7\\ \hline
Program 12	&8	&0	&13\\ \hline
Program 13	&6	&0	&16\\ \hline
Program 14	&1	&0	&22\\ \hline
Program 15	&2	&2	&18\\ \hline
Program 16	&14	&0	&0\\ \hline
Program 17	&1	&0	&26	\\ \hline
Total		&39	&6	&175\\ \hline
\end{tabular}
\end{table} 

\begin{table}[!t]
\renewcommand{\arraystretch}{1.3}
\caption{The number of parameter values of different sizes for every model after the reduction (VPP=Values per parameter).}
\label{table:param_values_reduction}
\centering
\begin{tabular}{|l|d{1}|d{1}|d{1}|}
\hline
Program&
VPP >99&
VPP (21-99)&
VPP <21 \\ \hline
Program 1	&0	&2	&4	\\ \hline
Program 2	&0	&0	&3	\\ \hline
Program 3	&0	&0	&4	\\ \hline
Program 4	&0	&0	&3	\\ \hline
Program 5	&0	&0	&4	\\ \hline
Program 6	&0	&0	&34	\\ \hline				
Program 7	&1	&0	&7	\\ \hline
Program 8	&0	&0	&7	\\ \hline
Program 9	&0	&2	&5	\\ \hline
Program 10	&1	&4	&3	\\ \hline
Program 11	&0	&0	&7	\\ \hline					
Program 12	&0	&8	&13	\\ \hline
Program 13	&0	&3	&19	\\ \hline
Program 14	&0	&1	&22	\\ \hline
Program 15	&1	&3	&18	\\ \hline
Program 16	&0	&14	&0	\\ \hline
Program 17	&0	&1	&26	\\ \hline
Total		&3	&38	&179\\ \hline
\end{tabular}
\end{table} 
Some of the input space models ended up having quite a large number of possible values per parameter (e.g, some parameters had up to 50000 values). In total, 39 parameters had more than 100 possible values assigned to each of them (more details are give in Table \ref{table:param_values_before_reduction}) and therefore we decided to reduce the number of possible parameter values for those parameters during the modeling process. The ACTS tool notifies the user modeling the system that the test generation might be slow when a system contains a parameter with more than 100 possible values. The reduction of the parameter values was conducted by asking a test engineer at Bombardier Transportation Sweden AB to reduce the possible parameter values. As a result of the reduction, only three parameters in three different programs contained one parameter with more than 100 possible values. Table \ref{table:param_values_reduction} shows the number of parameters with different values per parameter after the value reduction.


\begin{figure}[!ht]
\centering
\includegraphics[width=0.8\columnwidth]{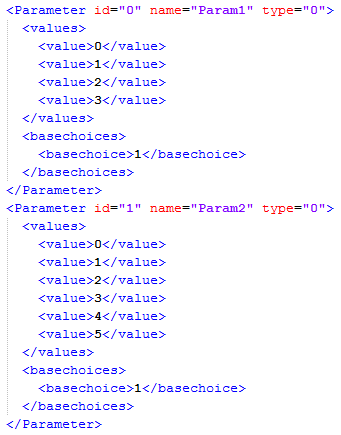}
\caption{A snippet showing two parameters from an XML representation of a system modeled in ACTS.}
\label{figure:ACTS_model_XML}
\end{figure}

To create a system for each input model in ACTS is a straightforward task since each data type in the model was directly converted to a corresponding type in ACTS. Figure \ref{figure:ACTS_model_XML} shows a snippet from a program modeled in ACTS, as an XML file.

\begin{mdframed}[style=style1]
{\it Answer RQ1: ACTS is applicable for modeling the input space of industrial PLC programs by using information directly contained in the source-code as well as information provided by engineers.}
\end{mdframed}

\subsection{Generation Time and Number of Test Cases}

\begin{table}[!t]
\renewcommand{\arraystretch}{1.3}
\caption{
The number of test suites generated under 600s for each program and algorithm with different interaction strengths. The percentage represents the number of test suites generated under 600s out of all the possible number of test suites for each algorithm.}
\label{table:less600s_sum}
\centering
\begin{tabular}{|l|l|l|l|l|}
\hline
 & IPOG& IPOG-F& IPOG-F2& IPOG-D \\ \hline
 Test Suites \textless 600s&	78&	60&	40&	50\\ \hline
Total Test Suites& 92&	92& 75& 75\\ \hline
Percentage	&85\%&	65\%&	53\%&	67\% \\ \hline
\end{tabular}
\end{table}

\begin{table}[!t]
\renewcommand{\arraystretch}{1.3}
\caption{The number of possible test suites generated under 600s for different program sizes.}
\label{table:less600s_per}
\centering
\begin{tabular}{|l|l|l|l|l|}
\hline
Program Size &IPOG	&IPOG-F	&IPOG-F2	&IPOG-D\\ \hline
Small&	100\%&	54\%&	70\%&	80\%\\ \hline
Medium&	93\%&	77\%&	60\%&	84\%\\ \hline
Large&	67\%&	47\%&	37\%&	43\%\\ \hline
\end{tabular}
\end{table}
In this section we report the results related to RQ2. We considered that a test suite generated in an efficient manner in terms of generation time if it has been created by ACTS in less than ten minutes according to information provided by the engineers working at Bombardier Transportation Sweden AB testing these programs. The information in Table \ref{table:less600s_sum} shows the number of modeled programs for which ACTS could generate a test suite in less than 600 seconds. The \textit{Total test suites} number represents the total number of different test suites that can be generated for each program (i.e., the number of test suites of all possible strengths of t allowed by the model and the algorithm). For example Program 3 contains four parameters. Since IPOG allows strength 1-6, for Program 3 we can only generate test suites up to strength 4. Therefore, it is possible to generate four different test suites using IPOG. Program 7 for example contains eight parameters, but we can only generate six test suites with different strengths using IPOG. In Table \ref{table:less600s_sum} we show the number of test suites generated under 600 seconds. The results in Table \ref{table:less600s_sum} show that the IPOG algorithm could generate 85\% of the possible test suites under 600s per modeled program and is therefore the most efficient in terms of generation time, followed by IPOG-D, IPOG-F, and IPOG-F2.

In Table \ref{table:less600s_per} we show for how many of the modeled programs ACTS can generate test suites using different algorithms in less than 600s for each category of program size. These results show that IPOG is the most efficient algorithm given the 600s time limit for generating the majority of test suites for all program sizes. For example, all test suites generated using IPOG for the programs considered small in size took less than 600s to generate.

\begin{table*}[!t]
\renewcommand{\arraystretch}{1.3}
\caption{Summary of the data collected for 2-way test cases. Generation time is given in seconds and size in number of test cases.}
\label{table:t2_data}
\centering
\begin{tabular}{|l|d{1}|d{2}|d{1}|d{2}|d{1}|d{2}|d{1}|d{2}|}
\hline
	&IPOG	&		&IPOG-F	&		&IPOG-F2&		&IPOG-D	& \\ \hline
	&Size	&Time	&Size	&Time	&Size	&Time	&Size		&Time 	\\ \hline
Avg	&1167	&0,049	&1162	&0,513	&1170	&1,156	&2428594	&5,915	 \\ \hline
Med	&231	&0,028	&231	&0,080	&231	&0,130	&3698		&0,014   \\ \hline
Min	&14		&0,014	&11		&0,013	&11		&0,003	&26			&0,003   \\ \hline
Max	&8828	&0,185	&8827	&5,086	&8827	&12,059	&25488036	&66,336  \\ \hline
\end{tabular}
\end{table*}

\begin{table*}[!t]
\renewcommand{\arraystretch}{1.3}
\caption{Summary of the data collected for 3-way test cases. Generation time is given in seconds, and size in number of test cases per test suite.}
\label{table:t3_data}
\centering
\begin{tabular}{|l|d{1}|d{2}|d{1}|d{2}|d{1}|d{2}|d{1}|d{2}|}
\hline
	&IPOG	&		&IPOG-F	&		&IPOG-F2&		&IPOG-D	& \\ \hline
	&Size	&Time	&Size	&Time	&Size	&Time	&Size		&Time 	\\ \hline
Avg&	5499	&0,109	&5446	&6,978	&5551	&13,038		&833331		&2,232 \\ \hline
Med&	210		&0,056	&210	&0,063	&211	&0,018		&350		&0,021  \\ \hline
Min&	28		&0,003	&28		&0,003	&28		&0,002		&42			&0,005  \\ \hline
Max&	20339	&0,560	&20339	&56,851	&20592	&118,307	&10707624	&28,569  \\ \hline
\end{tabular}
\end{table*}

\begin{table}[!t]
\renewcommand{\arraystretch}{1.3}
\caption{Results showing the generation time data for four programs that were not included in the overall evaluation for strength t=3. NA means that the data is not available.}
\label{table:t3_data_not_included}
\centering
\begin{tabular}{|l|l|l|l|l|}
\hline
	&IPOG	&IPOG-F		&IPOG-F2	&IPOG-D	\\ \hline
Program 10&	3.3s&	\textgreater3600.0s&	\textgreater3600.0s&	0.2s	\\ \hline
Program 12&	66.5s&	3554.6s&	\textgreater3600.0s&	69.5s      \\ \hline
Program 15&	23.2s&	\textgreater3600.0s&	\textgreater3600.0s&	NA      \\ \hline
Program 16&	118.9s&	\textgreater3600.0s&	2560.3s&	NA       \\ \hline
\end{tabular}
\end{table}

In Table \ref{table:t2_data} we show the average, median, minimum and maximum value for the generation time and test suite size, based on the results obtained from 15 programs when the interaction strength is set to two. No test data from Program 15 and 16 were included since the test generation ran out of memory using IPOG-D. In addition, in Table \ref{table:t3_data} we show the results for 3-way coverage. The data in this table is based on the data obtained from 13 out of the 17 programs. We were unable to use the rest of the programs (the non-included programs are shown in Table \ref{table:t3_data_not_included}) since only IPOG could be used for test generation within 600 seconds, with IPOG-F and IPOG-F2 performing very badly in terms of test generation efficiency. 

The data in Table \ref{table:t2_data} and Table \ref{table:t3_data} show that IPOG-F, on average, generates smaller test suites than IPOG, but at the cost of a longer generation time on average. Table \ref{table:t3_data_not_included} shows that IPOG-F has a longer generation time (i.e., longer than 1 hour generation time for 3 programs) than IPOG (i.e., less than 2 minutes generation time for the same 3 programs).

\begin{figure}
\centering
\includegraphics[width=1.05\columnwidth]{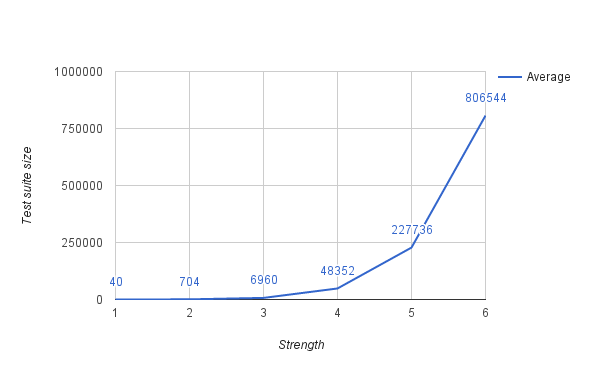}
\caption{The average test suite size using IPOG algorithm for different interaction strengths.}
\label{figure:test_size_avg}
\end{figure}

Our experiment shows that the size of a test suite increases with the interaction strength used. Figure \ref{figure:test_size_avg} shows the average test suite size using the IPOG algorithm for eight programs. The average is calculated using those programs where IPOG could generate a test suite for all strengths from 1 to 6. The plot shows the exponential increase in test suite size with the strength t.

\subsubsection{Results on IPOG-D}
According to the user guide of ACTS \cite{acts292userguide17} a moderate sized model contains less than 20 parameters and approximately 10 values per parameter. For these programs IPOG, IPOG-F and IPOG-F2 are preferred to be used for combinatorial test generation. For larger systems IPOG-D is suggested as a preferred way instead of the other algorithms.
Six of our models under test have more than 20 parameters (i.e., Program 6, 12-15 and 17 shown in Table \ref{table:param_values_avg}). Eight modeled programs have more than 10 values per parameter on average (i.e., Program 1, 7, 9, 10, 12 and 15-17). Three programs contain more than 20 parameters and more than 10 values per parameter on average (i.e., Program 12, 15 and 17). These three programs are considered large according to the ACTS official user guide. 

Our experiments show that when using the IPOG-D strategy Program 12 runs out of memory for strength 5, Program 15 runs out of memory when the strength is set to 2, and Program 17 reaches the cut-off time within one hour when the strength is 6. Analyzing the results from the other programs (i.e., which were defined as large PLC programs by engineers in Bombardier Transportation Sweden AB), showed that Program 13 reaches the cut-off generation time of one hour at strength 5, Program 14 creates all possible test suites, and Program 16 runs out of memory when the strength is set to 2. When using IPOG and IPOG-D for our definition of large programs, the results show a smaller generation time for IPOG-D in 4 out of 14 test suites, but at the cost of much larger test suites when compared to IPOG. In general IPOG-D creates larger test suites than the other algorithms for large programs.

\subsubsection{Base Choice Criterion}

\begin{table}[!t]
\renewcommand{\arraystretch}{1.3}
\caption{Generation times in seconds for different sizes of programs using the BaseChoice algorithm.}
\label{table:efficiency_time_BC}
\centering
\begin{tabular}{|l|d{2}|d{2}|d{2}|d{2}|}
\hline
\textbf{} & \multicolumn{4}{l|}{Generation Time} \\ \hline
\textbf{} & \textbf{Min} & \textbf{Max} & \textbf{Median} & \textbf{Average} \\ \hline
Small	&0,001	&0,002	&0,002	&0,002	\\ 
Medium	&0,002	&0,004	&0,002	&0,002	\\ 
Large	&0,002	&0,044	&0,003	&0,010	\\ \hline
Overall	&0,001	&0,044	&0,002	&0,005	\\ \hline
\end{tabular}
\end{table}

Generating tests using the BaseChoice algorithm results in a shorter generation time than pairwise testing (i.e., less than one tenth of a second, for all of the programs regardless of program size). More details are shown in Table \ref{table:efficiency_time_BC}) in which all measured generation times are measured in seconds. From our experiments, we show that ACTS using base choice criteria is efficient in terms of generation time regardless the size of the programs considered.

\subsubsection{Long generation times and running out of memory}

\begin{table}[!t]
\renewcommand{\arraystretch}{1.3}
\caption{Shows at which strength the algorithms have reached the cut time (3600s), or ran out of memory (OOM)  when applied to the programs. The 7 programs not included in the table could generate all of the possible test suites, as well as the combinations marked with "-".}
\label{table:OOM_3600s}
\centering
\begin{tabular}{|l|l|l|l|l|}
\hline
Program &	IPOG&	IPOG-F&	IPOG-F2&	IPOG-D \\ \hline
Program 1	&-			&5, \textgreater3600	&5, OOM		&6, \textgreater3600\\ \hline
Program 7	&-			&5, \textgreater3600	&4, OOM		&5, \textgreater3600\\ \hline
Program 9	&-			&-			&4, OOM		&-\\ \hline
Program 10	&5, OOM		&3, \textgreater3600	&3, \textgreater3600	&5, \textgreater3600\\ \hline
Program 12	&4, \textgreater3600	&4, \textgreater3600	&3, \textgreater3600	&5, OOM\\ \hline
Program 13	&5, \textgreater3600	&4, \textgreater3600	&4, \textgreater3600	&5, \textgreater3600\\ \hline
Program 14	&-			&6, \textgreater3600	&5, OOM		&-\\ \hline
Program 15	&4, \textgreater3600	&3, \textgreater3600	&3, \textgreater3600	&2, OOM\\ \hline
Program 16	&4, \textgreater3600	&3, \textgreater3600	&4, \textgreater3600	&2, OOM\\ \hline
Program 17	&-			&5, \textgreater3600	&5, OOM		&6, \textgreater3600\\ \hline
\end{tabular}
\end{table}

As previously mentioned, we used a cut-off time of one hour (3600 seconds) for test case generation. This means that if a test suite took longer than one hour to generate we stopped the process and no further tests with a higher strength were generated using the specific algorithm for the program under test. This occurred among the different algorithms and this occurrence differed from one program to another. For IPOG this stopping criteria happened four times, for IPOG-F eight times and for IPOG-F2 and IPOG-D five times. The ACTS application also ran out of memory when generating a number of test suites. For IPOG this happened once, for IPOG-F2 five times and for IPOG-D three times. Interestingly for IPOG-F this cut-off did not happen (as shown in Table \ref{table:OOM_3600s}). What needs to be taken into consideration in this case is the strength for which the cut-off stopping criteria occurred when comparing these algorithms, since no further tests with a higher strength were generated after the tool stopped its execution.

\begin{mdframed}[style=style1]
{\it Answer RQ2: ACTS is efficient in generating test cases using Base Choice regardless of the size of the PLC program, but is showing limitations in generating t-wise test cases for all interaction strengths.}
\end{mdframed}

\section{Discussion}

This study provides insights and experiences on how to apply combinatorial modeling and testing to PLC industrial control software. The experiment is an empirical exploration into the use of ACTS for this type of software. The modeling process shows how ACTS can be used for modeling several programs of different size and modeling complexity as well as generating combinatorial test suites. 

To answer RQ2 we generated tests for 17 systems created in ACTS, one system for each PLC program. All five algorithms, IPOG, IPOG-F, IPOG-F2, IPOG-D and Base Choice were applied with different strength for t-way testing when possible. Our experiments show how the generation time and number of tests differ when using different algorithms combined with varying strengths of t-way coverage. 
Our results suggest that ACTS is not able to generate test suites under the one hour cut-off time for all interaction strengths of the t-wise strategies and algorithms used. For several programs, different algorithms and t-strengths caused the ACTS tool to run out of memory before finishing the generation of a test suite. IPOG creates the most test suites under 10 minutes and the test generation did not run out of memory or exceeded the cut-off time as many times compared to some of the other strategies.

IPOG-D was hypothesized \cite{acts292userguide17} to be suited for larger programs, but the hypothesis is not supported by our results. More studies are needed to further confirm the results of this study and get a better understanding of the circumstances in which this hypothesis can be confirmed.

Related to our work, Lei et al. \cite{lei2007ipog} conducted a study measuring test suite size and generation time of FireEye (an earlier version of ACTS). FireEye uses a test generation algorithm for t-way testing based on the IPO strategy. Different configurations of parameters were used to perform experiments. Their experimental results show an exponential increase in number of tests with the strength t, which is similar to the results of this paper. As a consequence, the generation time and the cost of executing and checking the results of the tests are potentially increasing. Another study by Hagar et al. \cite{hagar2015introducing} describes the results of adopting combinatorial testing tools (i.e., mainly using ACTS) in several pilot projects at one company. Overall the adoption verdict was positive, but changing an already implemented testing process can be demanding on the company and the engineers using the ACTS tool. 

Our study can contribute to the adoption knowledge needed by a company in the embedded system and automation domain. Our case study shows insights on how to use combinatorial testing but also on the expectations of applying ACTS for programs of different size. In the study by Hagar et al. \cite{hagar2015introducing}, the authors use a combinatorial coverage tool to check the coverage of already existing test suites. Ideally, there is an already existing test suite that can be used to cover a certain t combination \cite{hagar2015introducing}. When generating tests for higher strengths of t-way testing, we observed that this can be rather expensive in terms of number of test cases and test generation time for larger programs. Therefore, one way to use ACTS would be to check the t-way coverage of existing manually created test cases and then use the option to verify the extent the test suite covers a certain t-way combinations \cite{acts292userguide17}.

Borazjany et al. \cite{borazjany2012combinatorial} applied ACTS to a real-world system (i.e., the ACTS tool itself). This study shows how to model the input space for generating tests. Their results show that the input space modeling is important and should be conducted thoroughly. Our results suggest that the input modeling is crucial since an engineer can use different sources of information for building a combinatorial model such as the code itself. An engineer using a combinatorial testing tool needs to consider the translation rules for different data types between the original program under test and the combinatorial model used for test generation. Our study suggests that the modeling of the input space cannot be directly performed by using only the information directly contained in the code. An engineer using ACTS would need to use the information provided by implementers and testers to obtain a usable model.

\section{Threats to Validity}

We use a cut-off point for test generation after interviews with engineers from Bombardier Transportation AB regarding the needed time for a tester to provide a set of tests for a desired program. Even if the efficiency are not easily generalizable to other systems, engineers concluded that one hour was a reasonable cut-off point for ACTS to terminate its search. Recall, however, that the aim of these experiments was not to
provide measures of test effectiveness in the sense of bug-finding, but instead to evaluate
the applicability of using ACTS for test generation and its success
in meeting its goals for the PLC programs used. We wanted to work with a realistic cut-off time that could be used in practice if this approach is to be adopted for testing PLC programs. Similarly, engineers considered the tool efficient if it generated test cases in less than 10 minutes. Therefore, we assumed an efficiency time limit of 10 minutes for the sake of this experiment. The choice of how to define this testing budget in practice is an open research question that is not in the scope of this study, but can affect the answers to the research questions.

The results are based on an experiment in one company in Sweden using 17 PLC programs. Even if this number can be considered small, we argue that having access to real industrial control software and the opportunity to collect information from engineers working in the embedded system domain can be representative. More case studies are needed to generalize these results to other systems and domains.

We performed interviews with several engineers working as professional developers and testers with a certain level of familiarity with the system under test. In addition, these engineers did not have prior knowledge of working with combinatorial modeling and test generation but they were provided with a short guideline on the experiment we performed.

We used the ACTS tool for combinatorial modeling and test generation. There are other tools for modeling and generating combinatorial tests and these may give different results. Nevertheless, the ACTS tool is based on well known algorithms and we assumed that its test suites are similar to the output produced by other combinatorial test input generation tools. 

\section{Conclusion}
Recently, researchers have shown an increased interest in the area of combinatorial testing that resulted in the development of a number of test generation tools. ACTS is one of the popular combinatorial tools that has been applied to several industrial systems. Nevertheless, there is a lack of evidence on how this approach can be applied to industrial control PLC software.

In this paper, we performed a systematic study to determine the use of combinatorial modeling and test generation for PLC software using ACTS. The results from modeling 17 programs provided by Bombardier Transportation Sweden AB show that the information needed to model the input space model can be created using different sources of information from both the PLC programs (directly) and from engineers testing these programs (indirectly). We created both abstract and concrete models depending on the data types used in the PLC programs.
The systems created in ACTS using these models were used for generating tests using different algorithms combined with different t-way strengths. Overall, the IPOG algorithm seems to yield the best results in terms of generating the most test suites under 10 minutes generation time per program for all possible strengths. According to our results, the IPOG-D algorithm is not particularly applicable to large programs as it was previously hypothesized. The results of this study can be used to find out areas which combinatorial test generation tools can be improved and hopefully our concrete insights will lead to future research.





%



\ifCLASSOPTIONcompsoc
  \section*{Acknowledgments}
\else
  \section*{Acknowledgment}
\fi

This work is partially funded from the Electronic Component Systems for European Leadership Joint Undertaking under grant agreement No. 737494 and The Swedish Innovation Agency, Vinnova (MegaM@Rt2).

\balance

\ifCLASSOPTIONcaptionsoff
  \newpage
\fi


\bibliographystyle{IEEEtran} 
\bibliography{ref.bib} 

%

%

\begin{IEEEbiography}{Michael Shell}
Biography text here.
\end{IEEEbiography}

\begin{IEEEbiographynophoto}{John Doe}
Biography text here.
\end{IEEEbiographynophoto}


\begin{IEEEbiographynophoto}{Jane Doe}
Biography text here.
\end{IEEEbiographynophoto}




\end{document}